# High Power Gamma-Ray Flash Generation in Ultra Intense Laser-Plasma Interaction

Tatsufumi Nakamura[a], James K. Koga[a], Timur Zh. Esirkepov[a], Masaki Kando[a], Georg Korn[b,c], Sergei V. Bulanov[a,d]

[a] Kansai Photon Science Institute, Japan Atomic Energy Agency, Kizugawa, Kyoto, Japan 6190215
[b] Max-Planck-Institut für Quantenoptik, Garching, Germany 85748
[c] Institute of Physics, Czech Academy of Sciences, Prague, Czech Republic 18221
[d] Prokhorov Institute of General Physics, Russian Academy of Sciences, Moscow, Russia 119991

**ABSTRACT**

When high-intensity laser interaction with matter enters the regime of dominated radiation reaction, the radiation losses open the way for producing short pulse high-power γ-ray flashes. The γ-ray pulse duration and divergence are determined by the laser pulse amplitude and by the plasma target density scale length. On the basis of theoretical analysis and particle-in-cell simulations with the radiation friction force incorporated, optimal conditions for generating a γ-ray flash with a tailored overcritical density target are found.

**Keywords:** PACS numbers: 12.20.-m, 52.27.Ep, 52.38. Ph

γ-rays have a broad range of applications in industry, material science, logistics for providing shipment security, medicine for sterilizing medical equipment and for treating some forms of cancer, e.g. gamma-knife surgery [1]. γ-rays from space provide insights into physical processes in distant astrophysical objects as exemplified by Gamma Ray Bursts, by cosmic ray acceleration at shock wave fronts, and by emission from pulsar environments, where γ-rays are generated via Bremsstrahlung, pion decay, inverse Compton scattering and synchrotron radiation of ultrarelativistic electrons rotating in magnetic fields (see [2]).

High energy photons are also emitted in the high-intensity laser light interaction with plasmas [3]. High efficiency γ-ray generation has been demonstrated in a number of experiments on the laser interaction with solid and gas targets where the main mechanism of their generation is the Bremsstrahlung radiation of fast electrons interacting with high Z-material targets [3, 4] (see also reviews [5] and references therein). In the present paper we show that the advent of the multipetawatt lasers can bring us into new regimes when the nonlinear Thomson scattering will produce the γ-ray flashes with extremely high efficiency of the laser energy conversion into the energy of γ-rays.

The characteristic energy of a photon emitted via nonlinear Thomson scattering, which has much in common with synchrotron radiation [6], scales with the electron quiver energy, $\gamma_e m_e c^2$, as $\mathcal{E}_\gamma \approx \hbar \omega \gamma_e^3$, where $\omega$ is the laser frequency and $m_e$ and $c$ are the electron mass and the speed of light in vacuum, respectively. The energy of the electron quivering in plasma under the action of an electromagnetic wave with an amplitude of $a = eE/m_e \omega c \gg 1$ is of the order of $m_e c^2 a$ [7]. For a laser frequency of the order of $10^{15} \mathrm{s}^{-1}$ the emitted photon energy is in the gamma-ray range if $a > 10^2$ which corresponds to a laser intensity higher than $10^{22} \mathrm{W/cm}^2$. The radiation generated by present-day lasers approaches this limit [8]. At this limit radiation friction effects change the electromagnetic wave interaction with matter rendering the electron dynamics dissipative, with efficient transformation of the laser energy into γ-ray photons.

Below the relativistic electron dynamics in the electromagnetic field is described by the equations of electron motion with the radiation friction force in the Landau-Lifshitz form [9]. When an electromagnetic wave propagates in an underdense plasma, or inside a self-focusing channel, its frequency $\omega$ and wave number $k$ (the wave vector component in the wave propagation direction) are related to each other through the dispersion equation $\omega^2 = k^2 c^2 + \Omega^2$. Here $\Omega$ is equal to $\omega_{pe}(1+a^2)^{-1/4}$ in the case of a circularly polarized plane wave [7] propagating in an underdense plasma with a density of $n_0$ corresponding to Langmuir frequency, $\omega_{pe} = (4\pi n_0 e^2 / m_e)^{1/2}$; it is equal to $1.84 c/R$ for the TE wave propagating inside a channel of radius $R$. The phase velocity of the wave, $v_{ph} = c\beta_{ph} = \omega/k$, is equal to $v_{ph} = c\omega/(\omega^2 - \Omega^2)^{1/2}$. The electron dynamics is considered in the boosted frame of reference where the electromagnetic wave is transformed into a spatially homogeneous electric field rotating with frequency $\Omega$. Retaining the main order terms in the radiation friction force in the Landau-Lifshitz form, one can write the electron equations of motion as

$$\dot{\mathbf{q}} = -\mathbf{a} - \frac{\varepsilon_{\mathrm{rad}}}{\gamma}\left\{\gamma^2 \dot{\mathbf{a}} - \mathbf{a}(\mathbf{q}\cdot\mathbf{a}) + \mathbf{q}\left[(\gamma \mathbf{a})^2 - (\mathbf{q}\cdot\mathbf{a})^2\right]\right\}, \quad (1)$$

where $\gamma = \left[1 + (p_1^2 + p_2^2 + p_3^2)/m_e^2 c^2\right]^{1/2}$ and the dot denotes differentiation with respect to the normalized time, $\tau$. Here we introduce normalized variables, $\tau = \Omega t$, $\mathbf{q} = \mathbf{p}/m_e c$, $\mathbf{a} = e\mathbf{E}/m_e \Omega c$. The dimensionless parameter $\varepsilon_{rad} = 2e^2\Omega/3m_e c^3$ determines the role of the radiation friction. The radiation friction effects become dominant when the laser pulse amplitude is equal to or greater than $a_{rad} = \varepsilon_{rad}^{-1/3}$ (see Ref. [10] and literature therein) corresponding for a one-micron wavelength laser to the intensity of $\approx 10^{23}\,\text{W/cm}^2$ with $a_{rad} \approx 400$, [11].

It is easy to show that the change of the momentum component parallel to the electromagnetic wave propagation, $q_1$, due to the radiation friction is negligible provided that the laser pulse duration is less than approximately 200 fs. The effect of the radiation friction on the rotating components of the electron momentum, $q_2, q_3$, is substantially stronger. In order to describe the electron motion we write the electron momentum as

$$\begin{pmatrix} q_1 \\ q_\| \\ q_\perp \end{pmatrix} = \begin{pmatrix} 1 & 0 & 0 \\ 0 & \cos(\tau) & \sin(\tau) \\ 0 & -\sin(\tau) & \cos(\tau) \end{pmatrix} \begin{pmatrix} q_1 \\ q_2 \\ q_3 \end{pmatrix}. \qquad (2)$$

Here $q_\|$ and $q_\perp$ are the components of the electron momentum parallel and perpendicular to the electric field, respectively. Substituting these expressions to equation (1) and neglecting the change of the $q_1$ – component, we obtain

$$\dot{q}_\perp - q_\| = -\varepsilon_{rad}\left[\gamma a + a^2 \frac{q_\perp}{\gamma}\left(1 + q_\perp^2\right)\right], \qquad (3)$$

$$\dot{q}_\| + q_\perp = a - \varepsilon_{rad} a^2 q_\| \frac{q_\perp^2}{\gamma}. \qquad (4)$$

Multiplying equation (3) by $u_\| = q_\|/\gamma$ and equation (4) by $u_\perp = q_\perp/\gamma$ and taking the sum we find

$$\dot{\gamma} = a u_\| - \varepsilon_{rad}\left(a q_\perp + a^2 q_\perp^2\right), \qquad (5)$$

which shows how the electron acquires the energy from the electromagnetic wave and loses it due to radiation friction.

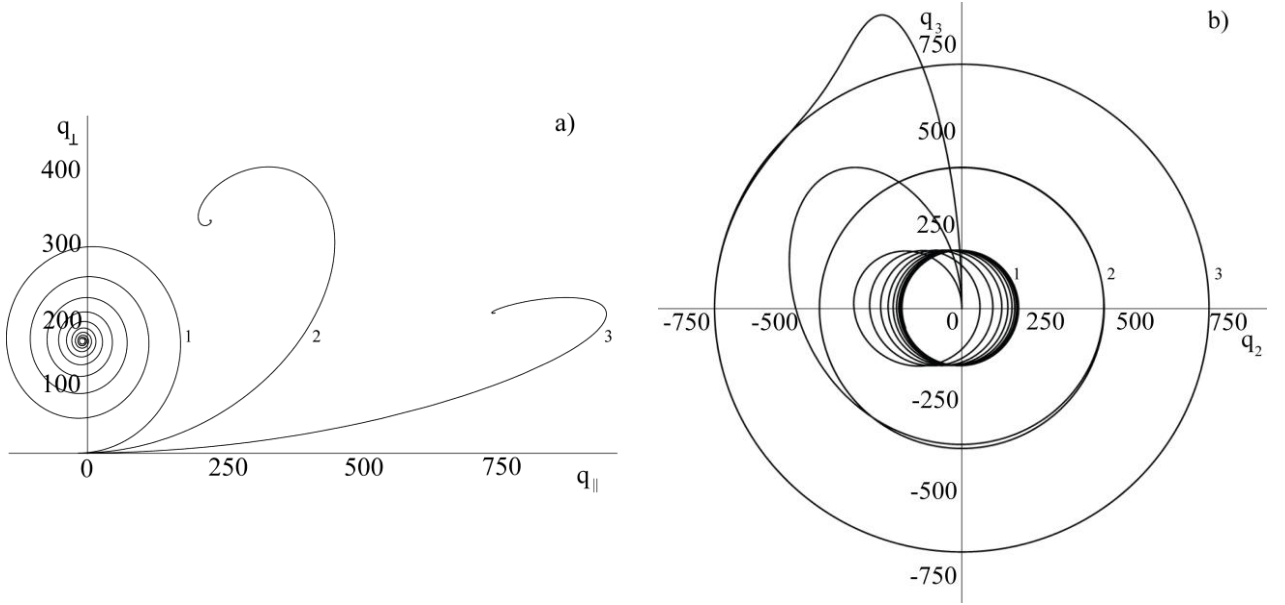

Figure 1. Electron orbits in a) the ($q_\|, q_\perp$) plane, and b) the ($q_2, q_3$) plane for $\varepsilon_{rad} = 10^{-8}$. Curves for (1) $a = 0.35\, a_{rad}$, (2) $a = a_{rad}$, and (3) $a = 5\, a_{rad}$

Typical solutions for this system of equations are presented in Fig. 1: where frames (a) and (b) show the electron orbits projections onto the ($q_\|, q_\perp$) plane and the ($q_2, q_3$) plane, respectively. As we see, at $a \ll a_{rad} = \varepsilon_{rad}^{-1/3}$ the electron oscillations in the ($q_\|, q_\perp$) plane decay slowly while for the laser pulse amplitude values equal to or above $a_{rad} = \varepsilon_{rad}^{-1/3}$, the electron oscillations in the rotating coordinate system decay during a time of the order of or less than the wave period. The asymptotics for $q_\|$ and $q_\perp$ are given by stationary solutions of the system of equations (3, 4). If the amplitude of electromagnetic wave is relatively small, i.e., $1 \ll a \ll a_{rad} = \varepsilon_{rad}^{-1/3}$, then for the components of the electron momentum perpendicular and parallel to the electric field Eqs. (3, 4) yield $q_\perp \approx a - \varepsilon_{rad}^2 a^7$, $q_\| \approx \varepsilon_{rad} a^4$. In the opposite limit, when $a \gg a_{rad} = \varepsilon_{rad}^{-1/3}$, we have $q_\perp \approx (\varepsilon_{rad} a)^{-1/2}$, $q_\| \approx (a/\varepsilon_{rad})^{1/4}$.

According to Eq. (5), the energy flux reemitted by the electron is equal to $e(\mathbf{v}\cdot\mathbf{E})$, which is $\approx \varepsilon_{rad} m_e c^2 \Omega \gamma (a q_\perp + a^2 q_\perp^2)$. The integral scattering cross section by definition [9] equals the ratio of the reemitted energy flux to the Poynting vector magnitude, $cE^2/4\pi$:

$$\sigma = \sigma_T \left( \frac{q_\perp}{a} + q_\perp^2 \right). \qquad (6)$$

Here $\sigma_T$ is the Thomson scattering cross section, $\sigma_T = 8\pi r_e^2/3 = 6.65\times 10^{-25}\text{cm}^2$. In the range of the wave amplitudes of $1 \ll a \ll a_{rad} = \varepsilon_{rad}^{-1/3}$, the integral scattering cross section grows as $\sigma \approx \sigma_T(1+a^2)$. It reaches a maximum of $\sigma \approx \sigma_T a_{rad}^2$ at $a = a_{rad}$, and then for $1 \ll a_{rad} \ll a$ it decreases according to $\sigma \approx \sigma_T a_{rad}^3/a^2$ .as seen in Fig. 2.

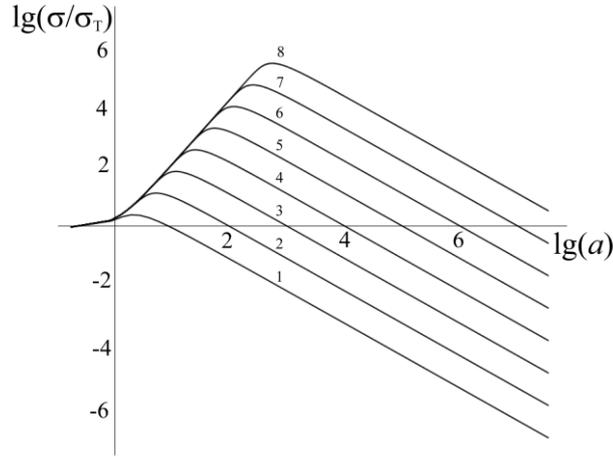

Figure 2. Dependence of $\lg(\sigma/\sigma_T)$ on $\lg(a)$. For each curve the integer label n corresponds to $\varepsilon_{rad} = 10^{-n}$.

The electron quivering in the laser field emits photons whose energy is proportional to the cube of the electron Lorentz factor: $\hbar\Omega_\gamma = \hbar\Omega\gamma^3$. For $1 \ll a \ll a_{rad} = \varepsilon_{rad}^{-1/3}$ a typical value of the photon frequency is proportional to $a^3$. In the limit of high laser intensity $1 \ll a_{rad} \ll a$ the frequency scales as $\Omega_\gamma = \Omega(a/\varepsilon_{rad})^{1/4}$.

The laser pulse depletion length is of the order of $l_{dep} \approx 1/\sigma n_e$. It reaches its minimum for given electron density at $a = a_{rad}$ when the integral scattering cross section is maximal:

$$\min\{l_{dep}\} \approx \frac{1}{\sigma_{max} n_e} = \frac{1}{\sigma_T n_e}\left(\frac{2r_e\Omega}{3c}\right)^{2/3}. \qquad (7)$$

For $\Omega \approx \omega$ with $\omega$ corresponding to a one-micron wavelength laser, the maximal value of the integral scattering cross section is of the order of $10^{-19}\text{cm}^2$ provided that the laser intensity is as large as $10^{23}\text{W/cm}^2$. If the laser with this intensity irradiates a solid density target ($\approx 10^{23}\text{cm}^{-3}$), then Eq. (7) gives $\min\{l_{dep}\}$ in the micron range. This results in a γ-ray flash with the duration and power comparable, within an order of magnitude, to the incident laser pulse duration and power.

In order to estimate the laser power required for realization of the optimal conditions for the gamma-flash emission we use a relationship between the laser power $\mathcal{P}_{las}$ and amplitude under the conditions of relativistic self-focusing found in Ref. [12]. It reads $a^3 = 8\pi \left( \mathcal{P}_{las} / \mathcal{P}_c \right) \left( \omega_{pe} / \omega \right)^2$ with $\mathcal{P}_c = 2 m_e^2 c^5 / e^2 \approx 17$ GW. The optimal condition, $a^3 = \varepsilon_{rad}^{-1}$, yields $\mathcal{P}_{las} \approx 10^2 \left( \omega / \omega_{pe} \right)^2$ PW, i.e., in the case of the target plasma density of the order of $10 n_{cr}$ the required laser power is about 10 PW.

Performing the Lorentz transform to the laboratory frame of reference we find that in the limit of strong radiation losses when $1 \ll a_{rad} \ll a$, we have $p_1 \approx m_e c \left( a / \varepsilon_{rad} \right)^{1/4} / \left( \beta_{ph}^2 - 1 \right)$, $p_\parallel \approx m_e c \left( a / \varepsilon_{rad} \right)^{1/4}$ $p_\perp \approx m_e c / \left( \varepsilon_{rad} a \right)^{1/2}$. As we see the radiating electrons move in the direction of the laser pulse propagation. This results in γ-photon energy upshifting by a factor $2 \left( \beta_{ph}^2 - 1 \right)^{-1/2}$ and to a γ-beam collimation within the angle $\left( \beta_{ph}^2 - 1 \right)^{1/2}$.

During interaction of super-high-power laser light with matter the laser pulse is a subject of various instabilities. Among them the most important is the relativistic self-focusing resulting in the laser pulse channeling, which is also called the "hole boring" in respect to the interaction with overdense targets. It leads to the increase of the laser pulse amplitude and to the decrease of the electron density in the interaction region, which change the laser energy depletion length and the parameters of the γ-rays emitted. Thorough studying of these effects and of the effects of the plasma inhomogeneity requires computer simulations.

We performed parametric studies of the laser pulse interaction with high density targets using the two-dimensional (2D) particle-in-cell (PIC) code [13] where the radiation friction force has been incorporated in the Landau-Lifshitz form as has also been done in Ref. [14].

In simulations, the laser pulse has the normalized amplitude of $a = 150$, a power of $\mathcal{P}_{las} = 10$ PW, an energy of 300 J, and a pulse duration of 30 fs, polarized in the y-direction. In the tailored plasma target the density changes from $0.1 n_c$ to $350 n_c$ exponentially, $n(x) \propto \exp(x/L)$, with the plasma inhomogeneity scale length, $L$, in the range from 0.1μm to 20μm, and then becomes constant having a thickness of 10 μm. The simulation box has the width equal to 80μm and the length varying from 50μm to 210μm. The mesh has a spatial resolution of $\Delta x = \Delta y$ varying from 1/40μm to 1/200μm with a temporal resolution of $\Delta t = 0.0025$ fs. The plasma is comprised of electrons and ions with a mass number to charge ratio corresponding to $A/Z = 2$. The number of particles of each species per cell is 50. Simulation results for the parameters of interest are shown in Fig. 3. The laser pulse interacts with the plasma target, whose density inhomogeneity is characterized by a scale length equal to L=2.5μm. Fig. 3 a) shows the dependence of the radiation power, $\mathcal{P}_\gamma$, and energy, $\mathcal{E}_\gamma$, on time. We see that the emitted γ-ray flash has a duration approximately equal to 30 fs with a maximal power equal to $\mathcal{P}_\gamma$=2.75 PW. The laser energy converted to the γ-rays is about 96 J corresponding to 32% efficiency. The laser pulse undergoes self-focusing and becomes confined inside the self-focusing channel which leads to an almost complete laser pulse energy absorption. The angular distribution of the emitted radiation has been calculated according to the formula [9]

$$dI = \frac{e^2}{4\pi c^3}\left\{\frac{2(\mathbf{n}\cdot\mathbf{w})(\mathbf{v}\cdot\mathbf{w})}{c(1-(\mathbf{v}\cdot\mathbf{n})/c)^5} + \frac{\mathbf{w}^2}{(1-(\mathbf{v}\cdot\mathbf{n})/c)^4} - \frac{(1-v^2/c^2)(\mathbf{n}\cdot\mathbf{w})^2}{(1-(\mathbf{v}\cdot\mathbf{n})/c)^6}\right\}do, \qquad (8)$$

where $\mathbf{n}$ is the unit vector in the emission direction, $\mathbf{v}$ and $\mathbf{w}=\dot{\mathbf{v}}$ are the electron velocity and acceleration, and $do$ is the element of solid angle. The summation was performed over all radiating electrons. Fig. 3 c) shows less divergence of the gamma-ray flash for a lower density plasma, where the laser pulse group velocity is larger, in agreement with above discussed the γ-ray beam collimation due to relativistic motion of the source with $\left(\beta_{ph}^2-1\right)^{1/2} \approx 0.5$ in our case. Two lobes seen in the gamma-ray angular distribution in Fig. 3 c, show that in the frame of reference co-moving with the gamma-ray source the emitting electrons have their transverse momentum component larger than the longitudinal.

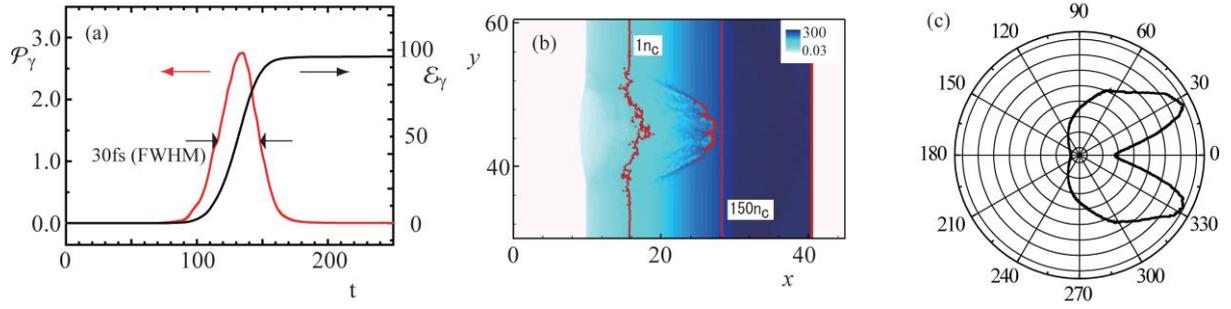

Figure 3. a) The radiation power, $\mathcal{P}_\gamma$ (PW), and energy, $\mathcal{E}_\gamma$ (J), vs time t (fs). b) The ion density distribution in the (x,y) plane for $t = 260\,\text{fs}$, constant density levels are shown for $n = n_c$ and $n = 150 n_c$. c) The γ-ray intensity angular distribution.

Our simulations reveal the dependences of the emitted γ-ray pulse energy, duration (it is the radiative loss time scale in the considered case of near critical plasma density) and power on the plasma density scale length, presented in Fig. 4. As we see in Fig. 4 a), the radiated γ-ray pulse energy increases when the scale length increases saturating at approximately $120\,\text{J}$ for $L = 15\,\mu\text{m}$. The gamma-ray pulse duration grows monotonously, Fig. 4 b). For the longer scale length, the laser energy depletion and resultant gamma-ray emission take place gradually, leading to less power emitted. In the case of relatively small scale length, the laser pulse reflects from the target with weak absorption, which results in a weaker γ-ray flash with its duration of the order of that of the laser pulse. Therefore we have an optimum plasma scale length for the high power gamma-ray flash emission, as seen from Fig.4 c). For chosen simulation parameters the radiated power reaches its maximum $\approx 2.75\,\text{PW}$ at $L \approx 2.5\,\mu\text{m}$.

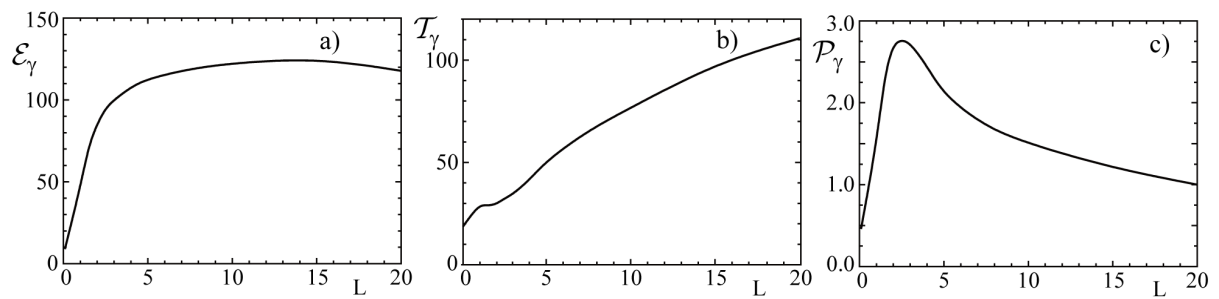

Figure 4. Dependences of the emitted γ-ray pulse energy, $\mathcal{E}_\gamma$ (J) a), duration, $\mathcal{T}_\gamma$ (fs) b) and power, $\mathcal{P}_\gamma$ (PW) c) on the plasma density scale length, L (μm).

For a fixed laser pulse energy and varying duration, i.e., varying power, the maximum γ-ray pulse power is reached at different scale lengths as shown in Fig. 5, corresponding to the laser pulse energy of 300 J. The found optimal plasma scale lengths are 1.2, 2.5, and 7.0 μm for the laser pulse length of 15, 30, and 60 fs, respectively, which are roughly the same as the pulse length.

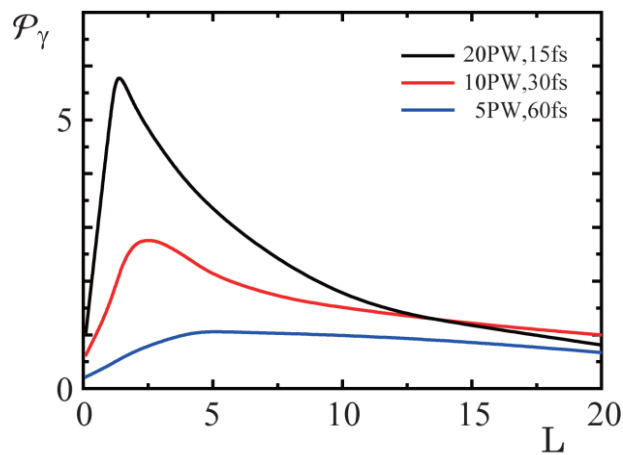

Figure 5. (a) Dependence of the γ-ray power $\mathcal{P}_\gamma$ (PW) on the plasma scale length, L (μm) for the laser pulse energy of 300 J and the laser power, $\mathcal{P}_{las}$, varying from 5 to 20 PW. (b) Energy conversion efficiency from laser to electron and radiation.

In conclusion, we show that in an ultra intense laser-plasma interaction, almost all the laser pulse energy can be converted into a strongly collimated γ-ray flash with almost the same power as the laser pulse. The γ-ray flash is generated due to the dominating role of the radiation friction force which completely transfigures the laser-matter interaction as is has been discussed in Refs. [15]. On the basis of theoretical analysis and particle-in-cell simulations with the radiation friction force incorporated, we found the optimal conditions for generating a γ-ray flash in the laser interaction with a tailored overcritical density target, for which the laser pulse to the gamma-ray energy conversion efficiency is substantially high. A one-micron wavelength laser pulse with the power of 10 PW can be converted into a gamma ray flash with the efficiency of about 32 %. We note that for a 1 PW, 30 fs laser pulse the conversion efficiency can be approximately 3% with a γ-ray flash duration of 30 fs. For the considered laser and target parameters the laser does not show the multidirectional hole boring observed in Ref. [16], which could be caused by the hosing instability [17], the Weibel instability of laser accelerated electrons [18] or by not optimal laser-plasma matching [19]. In addition, as it has been demonstrated in Ref. [13] the radiation friction leads to the self-focusing patterns with fewer filaments than in the case without radiation friction effects taken into account.

The realization of these regimes will be feasible with the next generation of high power lasers [20]. The proposed source of short pulse high-power strongly collimated γ-rays will benefit fundamental and applied sciences and laboratory astrophysics [21]. In particular, in studying photo-nuclear reactions [4, 5], copious production of electron-positron pairs [22] and in studying of photon-photon collisions [23] in the low energy range.

**Acknowledgments**

The authors thank A. Ya. Faenov, J.-P. Contzen, L. A. Gizzi, P. Kaw, A. Macchi, F. Pegoraro, T. A. Pikuz, A. S. Pirozhkov, H. Ruhl, M. Tamburini, and A. G. Zhidkov for discussions. The simulations are performed by using the Primergy BX900 at JAEA-Tokai.